\documentclass{ifacconf}

\usepackage{graphicx}      % include this line if your document contains figures
\usepackage{natbib}        % required for bibliography
\usepackage[english]{babel}

\begin{document}
\begin{frontmatter}

\title{Truncated L\'evy statistics for\\
transport in disordered semiconductors\thanksref{footnoteinfo}}

\thanks[footnoteinfo]{Authors are grateful to the Russian
Foundation for Basic Research (grant~10-01-00608) for financial
support.}

\author[First]{Renat T. Sibatov},
\author[Second]{Vladimir V. Uchaikin}

\address[First]{Ulyanovsk State University,
   Russia (e-mail: ren\_sib@bk.ru)}
\address[Second]{Ulyanovsk State University,
   Russia (e-mail: vuchaikin@gmail.com)}

\begin{abstract}\vspace{0.3cm}

Probabilistic interpretation of transition from the dispersive
transport regime to the quasi-Gaussian one in disordered
semiconductors is given in terms of truncated L\'evy
distributions. Corresponding transport equations with fractional
order derivatives are derived. We discuss physical causes leading
to truncated waiting time distributions in the process and
describe influence of truncation on carrier packet form, transient
current curves and frequency dependence of conductivity.
Theoretical results are in a good agreement with experimental
facts.
\end{abstract}

\begin{keyword}
truncated L\'evy distribution, disordered semiconductor, anomalous
transport, fractional derivative
\end{keyword}

\end{frontmatter}

%===============================================================================

\section{Introduction}

Various transport regimes can be observed in disordered
semiconductors: normal regime characterized by Gaussian statistics
and standard diffusion equation, and different types of anomalous
(non-Gaussian) transport. Among anomalous transport regimes, the
dispersive transport is distinguished especially (\cite{Sch:75,
Zvy:84, Mad:88}). The dispersive diffusion packet has a
non-Gaussian form, but even so it maintains its shape and its
spatial extent depends on time. In other words, the process
reveals the property of self-similarity (\cite{Sch:75}).
Dispersive transient current curves are sufficiently differ from
the normal ones corresponding to Gaussian transport. Current
decays according to very stretched law, two power law sections are
picked out: $I(t)\propto t^{-1+\alpha}$ for $t<t_{\rm T}$, and
$I(t)\propto t^{-1-\alpha}$ for $t>t_{\rm T}$. The parameter
$0<\alpha<1$ called the dispersion parameter depends on material
structure and transport mechanism. For some mechanisms,
temperature dependence is observed~(\cite{Zvy:84}). \cite{Sch:75}
have broken through in understanding of probability causes of
dispersive transport, $\alpha$ is interpreted as an exponent of
power law asymptotes in a distribution of sojourn times $T$ in
localized states (waiting times). Since $\alpha<1$, the mean value
$\langle\tau\rangle$ diverges. This leads to non-decreasing
relative fluctuations of the number of jumps between localized
states. As a consequence we have non-Gaussian form of a diffusion
packet and anomalous law of packet widening ($\Delta\propto
t^{\alpha/2}$).

The known approaches (such as the Scher-Montroll model~(1975), the
Arkhipov-Rudenko theory~(1982, 1983)) connect the anomalous-normal
transition in the same material with increase of $\alpha$ to 1
because of changing outer conditions (for example, temperature).
Here, we consider another case, the transition occurs due to
changes of scale parameters. For small transient times, i.~e.
small values of sample thickness and/or large values of voltages,
normalized transient current curves are practically universal and
these curves correspond to dispersive transport. But in samples
with larger thickness, or at smaller voltages (at the same
temperatures), transient current curves have
plateau~(\cite{Bas:93, Tyu:05}), that is typical for Gaussian
transport. Results in this case are often described in frames of
the quasi-equilibrium transport theory. In other words, at
increasing transient times, the tendency towards the
quasi-Gaussian statistics is observed. This tendency can not be
explained by change of the dispersive parameter $\alpha$, it is a
spatiotemporal scale effect.

Electron transport in polymers is usually modelled as a hopping
process. Experimental time-of-flight results for polymers come to
an agreement with the theory, when the Gaussian form of an energy
distribution of localized states is taken~(\cite{Bas:93}).
Transition from dispersive type of transport to the quasi-Gaussian
one can be explained by such form of density of localized states.

The goal of the present paper is to give a probabilistic
interpretation of transition from the dispersive transport regime
to the quasi-Gaussian one due to change of the scale parameters
characterizing the process. The additional question is to
determine main features of energetic and/or spatial distributions
of traps leading to the discussed phenomenon. Truncated L\'evy
distributions introduced by \cite{Man:94} play an important role
in our interpretation. We derive correspondent equations with
fractional order derivatives and discuss the influence of
truncation on transient current curves and frequency dependence of
conductivity.

\section{Dispersive transport and fractional differential approach}

Dispersive (non-Gaussian) transport~(\cite{Mad:88}, \cite{Zvy:84})
is observed in many disordered materials differ in its
microscopical structure. A comparison of available data suggests
the presence of universal transport properties unrelated to the
detailed atomic and molecular structure of a substance. The
fractional differential approach is often applied to description
of dispersive transport (\cite{Bar:01, Uch:05, Uch:08, Sib:09}).

The Riemann-Liouville type fractional derivatives
$$
_0\textsf{D}_t^\alpha
f(t)=\frac{1}{\Gamma(1-\alpha)}\frac{d}{dt}\int\limits_0^t
\frac{f(t')dt'}{(t-t')^\alpha}, \quad 0<\alpha<1
$$
were for the first time applied to description of dispersive
transport by \cite{Ark:83b}. The authors expressed the
relationship between concentrations of free and localized carriers
through the fractional integral. In later papers they use a
different approximate relation between concentrations of localized
and free carriers, which they called the master dispersive
transport equation. This relation is believed to hold for any
density of localized states and permits expressing results through
elementary functions in the case of an exponential density. The
Arkhipov-Rudenko master equation leads to a diffusion equation
with a variable diffusion coefficient and
mobility~(\cite{Ark:83}).

On the base of the kinetic trapping-emission equations,
Tiedje~(1984) derived a transport equation for free carrier
concentration. The inverse Laplace transform of this equation is
nothing but a fractional differential equation~(\cite{Sib:07}).

\cite{Bar:01} made use of the fractional Fokker-Planck equation to
account for transient photocurrent relaxation in amorphous
semiconductors. He showed the agreement between selected results
of the fractional differential approach and results predicted by
the Scher-Montroll model~(1975).

Power-law decay of photoluminescence in amorphous semiconductors
was described in Refs~(\cite{Sek:03, Sek:06}) in frame of the
generalized random walk model with recombination by tunnel
radiative transitions. Recombination was limited by dispersive
diffusion of the carriers. In the framework of this model,
\cite{Sek:06} compiled a fractional differential equation for the
first passage time distribution density. The recombination rate
was found using the integral Laplace transform of this equation.

As shown in Refs~(\cite{Uch:99b, Sib:07}), the main asymptotic
terms in the Scher-Montroll model satisfy fractional differential
equations, the Green functions of which are fractionally stable
densities.

\section{The Scher-Montroll model}

Continuous time random walk (CTRW) model, introduced by
\cite{Sch:75}, provided the first detailed explanation of all the
main patterns of current behavior in time-of-flight experiments
with amorphous semiconductors.

The main assumptions of this model are as follows:
\begin{enumerate}

\item The transport of charge carriers is a jumplike random walk
in which the walkers change their positions at random instants of
time.

 \item Carrier jumps are independent of one another, and
time intervals between them (waiting times) are independent,
identically distributed random variables~$T$.

 \item Waiting times are characterized by asymptotically power-law
distribution:
\begin{equation}\label{power}
\textsf{P}\{T>t\}\propto t^{-\alpha},\quad t\rightarrow\infty,
\qquad 0<\alpha<1.
\end{equation}

\end{enumerate}

\cite{Sch:75} simulated charge transfer in disordered
semiconductors as carrier hopping within the model grid of
localized states. The grid constitutes a regular cubic lattice,
each cell of which contains randomly distributed sites (localized
states). The waiting time till the next hopping depends on the
distance to the nearest neighbor sites. The cell residence time
distribution can obey the power law owing to site spatial disorder
in a cell.

As is known, the description of normal transport is based on the
central limit theorem of the probability theory. For random
quantities distributed according to asymptotically power law,
divergence of dispersion for ${\alpha<2}$ and divergence of
mathematical expectation for $\alpha<1$ make this theorem
inapplicable, which necessitates the application of the
generalized limit theorem.

\

\textbf{The generalized limit theorem.} Let random quantities
$X_j$ be independent and identically distributed, and satisfy the
following conditions
$$
{\sf P}(X > x) \sim a_+ x^{-\alpha}, \qquad x \to \infty,
$$
$$
{\sf P}(X < - x) \sim a_- x^{-\alpha}, \qquad x \to \infty,
$$
$0 < \alpha \leq 2, \ a_+ \geq 0, a_- \geq 0$ and $a_+ + a_- > 0$.
Then, $A_n$ and $B_n > 0$, sequences exist such that, as $n \to
\infty$, one finds
$$
\left. \left( \sum_{j=1}^{n} X_j - A_n \right)\right/ B_n \mathop
{ \sim }\limits^{d} S^{(\alpha,\beta)},
$$
where $S^{(\alpha,\beta)}$ is the stable random variable with
exponent $\alpha$ and asymmetry parameter ${\beta = (a_+ -
a_-)/(a_+ + a_-)}$. Stable random quantities can be defined
through characteristic functions having the form (form A)
(\cite{Uch:99a}):
$$
g^{(\alpha,\beta)}(k)=\exp\{-|k|^\alpha[1-i\beta\tan(\alpha\pi/2)\
\mathrm{sign}(k)]\},\qquad \alpha\neq 1,
$$
$$
g^{(1,\beta)}(k)=\exp(-|k|).
$$

Certainly, there are an infinite number of sequences of
normalizing coefficients $A_n, B_n$ showing similar asymptotic
behavior as $n \to \infty$. By way of example, they can be defined
in the following way ($a = \langle X \rangle$ and $c = a_+ +
a_-$): \ \vspace{0.3cm}\\ {\small
\begin{tabular}{lll}
при $\alpha = 2$& $A_n=na,$&$B_n = \sqrt{cn\ln n}$,\\
при $\alpha \in (1,2)$& $A_n=na,$&$
B_n=\left(\pi cn/[2\Gamma(\alpha)\sin(\alpha\pi/2)]\right)^{1/\alpha}$,\\
при $\alpha = 1$& $A_n = \beta c n \ln n,$&$ B_n = \pi cn/2$,\\
при $\alpha \in (0, 1)$&$A_n = 0,$&$
B_n=\left(\pi cn/[2\Gamma(\alpha)\sin(\alpha\pi/2)]\right)^{1/\alpha}$.\\
\end{tabular}
} \

In the Scher-Montroll model, waiting times $T$ (positive random
quantities) are distributed according to an asymptotically power
law. Therefore, at macroscopic scales, the first passage time
distribution, conduction current density, and concentration of
delocalized carriers considered as functions of time should have
the form of stable density distribution.

The main characteristics of the CTRW model are waiting time and
jump vector distributions, $\psi(t)$ and $p(\mathbf{r})$,
respectively. Spatial distribution density $p(\mathbf{r},t)$ of a
particle executing random walks and initially located at the
origin of coordinates is defined in terms of the Fourier-Laplace
transform~(\cite{Mon:65})
\begin{equation}\label{eq_CTRW_distribution}
\widehat{\widetilde{p}}(\mathbf{k},s)=\frac{1-\widetilde{\psi}(s)}{s}\frac{1}{1-\widehat{p}(\mathbf{k})\widetilde{\psi}(s)},
\end{equation}
where
$$
\widehat{\widetilde{p}}(\mathbf{k},s)=\int\limits_\mathbf{R}
d\mathbf{r}\ e^{i\mathbf{k}\mathbf{r}}\int\limits_0^\infty dt\
e^{-s t}p(\mathbf{r},t)
$$
is the Fourier-Laplace transform of normalized particle
concentration, $\widetilde{p}(\mathbf{k})$ is the Fourier
transform of path distribution density, and $\widehat{\psi}(s)$ is
the Laplace transform of waiting time distribution density.
Substituting into Eq.~(\ref{eq_CTRW_distribution}) the asymptotic
series expansion of the Laplace image of waiting time distribution
density with the power-law tail
$$
\widehat{\psi}(s)\sim 1-s^\alpha/c^\alpha,\quad s\ll c,
$$
along with asymptotic expansion of the Fourier image of path
distribution density:
$$
\widetilde{p}(\mathbf{k})\sim 1+\mathbf{c}_1 i \mathbf{k}-c_2 k^2,
\quad
  \ |\mathbf{k}|\ll 1/|\mathbf{c}_1|,
$$
and applying the Tauberian theorem, we obtain
$$
\widehat{\widetilde{p}}(\mathbf{k},s)=\frac{c^{-\alpha}
s^{\alpha-1}}{-\mathbf{c}_1 i \mathbf{k} + c_2 k^2 +
s^\alpha/c^\alpha}.
$$
Rewriting the last expression in the form
$$
[s^\alpha-\mathbf{c}_1 c^\alpha i \mathbf{k} + c_2 c^\alpha k^2 ]\
\widehat{\widetilde{p}}(\mathbf{k},s)=s^{\alpha-1}
$$
and applying inverse Fourier and Laplace transformations yield
\begin{equation}\label{eq_fractional_ScherMontroll}
_0\textsf{D}^\alpha_t p(\mathbf{r},t)+\mathbf{K}\nabla
p(\mathbf{r},t) - C \nabla^2
p(\mathbf{r},t)=\frac{t^{-\alpha}}{\Gamma(1-\alpha)}\delta(\mathbf{r}),
\end{equation}
where $\mathbf{K}$ and $C$ are vectorial and scalar constants.

\section{Truncated waiting time distributions}

It has been emphasized above that the self-similar dispersive
transport in disordered semiconductors is characterized by
asymptotically power law distributions of sojourn times $T$ of
carriers (electrons and/or holes) in localized states:
$\textsf{P}\{T>t\}\propto t^{-\alpha}$, $t\rightarrow\infty$,
where $0<\alpha<1$ is the dispersion parameter. Mean value of such
random variables diverges.

It is naturally to suppose that an asymptotically power law
distribution of waiting times can be truncated. This truncation
can be caused by finite values of mobility gap at multiple
trapping or by limitation of jump lengths at hopping. Secondary
mechanism acting in parallel to the main transport mechanism can
be responsible for the truncation. We shall consider an influence
of truncation of power law distributions of waiting times on
properties of dispersive transport. This influence should become
apparent in scale effects.

Mantegna and Stanley (1994) introduced truncated L\'evy flights, a
process showing a slow convergence to a Gaussian. The truncated
L\'evy flight is a Markovian jump process, with the length of
jumps showing a power-law behavior up to some large scale. At
large scales distribution has cutoffs and thus have finite moments
of any order. Smoothly (exponentially) truncated L\'evy flights,
introduced by Koponen (1995), constructed on Mantegna and
Stanley's ideas, allows to give a convenient analytic
representation of results.

In our model, jump lengths are distributed exponentially and
waiting times have asymptotically truncated power law
distributions. We take this distribution in the form
\begin{equation}\label{trunc}
\Psi(t)=\textsf{P}\{T>t\}\sim \frac{(c
t)^{-\alpha}}{\Gamma(1-\alpha)} \exp(-\gamma t),
\end{equation}
$$
\quad \gamma\ll c,\qquad t\gg c^{-1}.
$$

\begin{figure}[tb]
\centering
\includegraphics[width=0.4\textwidth]{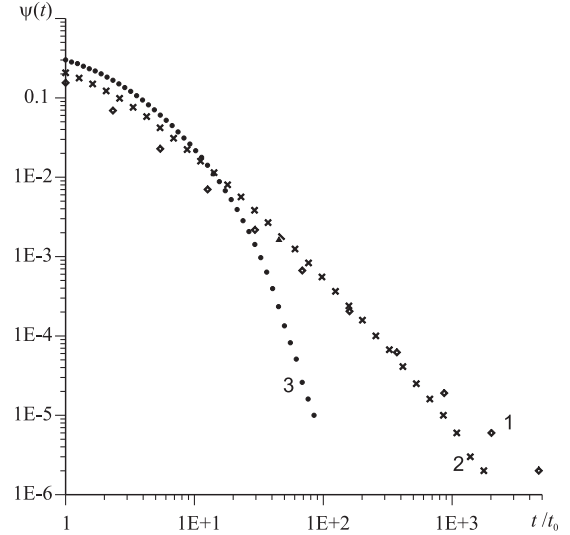}
\caption{Waiting time PDF's for different types of densities of
localized states in the multiple trapping model: exponential (1),
Gaussian (2) and rectangular~(3).}\label{fig_WT}
\end{figure}

Since the "heavy"\ power law tail is truncated, all moments of
random variable are finite, and consequently, a distribution of a
sum of large number of such random variables tends to the Gaussian
law. However, this convergence is very slow and stable L\'evy
distributions play a role of intermediate asymptotics.

In Fig.~\ref{fig_WT}, normalized waiting time distributions
numerically calculated for three forms of density of states
(exponential, Gaussian and rectangular) in the multiple trapping
model are presented. For all three cases, an average energy of
states $\varepsilon_0$ is the same. For exponential DOS waiting
times are distributes according to the asymptotical power law with
the exponent $\alpha=\varepsilon_0/kT$. For the case of the
Gaussian DOS, sojourn times in traps are distributed according to
a wide law but all moments finite. For the case of rectanguler
DOS, waiting times have distribution that can be approximated by a
stretched exponential law.

Transport is dispersive in all time scales only in the case of
ideal exponential energy distribution of localized states. For two
other DOS, transport takes features of the normal transport in
asymptotics of large times.

\section{Transport equation for the case of truncated waiting time
distributions}

Connection between concentrations of delocalized (conduction)
$n_\mathrm{c}$ and trapped $n_\mathrm{t}$ carriers is expressed by
the formula
$$
n_\mathrm{t}(\mathbf{r},t)=\int\limits_0^t \tau_0^{-1}
n_\mathrm{c}(\mathbf{r},t')\ \Psi(t-t')\ d t',
$$
where $\tau_0$ is a mean time of delocalized state. Passing on to
Laplace transforms, taking into account the relation
$$
\hat{\Psi}(s)=\int\limits_0^\infty  e^{-st}\Psi(t)\ dt\sim
c^{-\alpha}(s+\gamma)^{\alpha-1},\quad s\ll c,
$$
we obtain $\hat{n}_\mathrm{t}(\mathbf{r},s)\sim
c^{-\alpha}\tau_0^{-1}(s+\gamma)^{\alpha-1}\hat{n}_\mathrm{c}(\mathbf{r},s)$.
The inverse Laplace transformation leads to the equation with
derivative of fractional order:
\begin{equation}\label{eq_relation}
n_\mathrm{c}(\mathbf{r},t)\sim c^\alpha\tau_0\ e^{-\gamma t}
{_0\textsf{D}_t^{1-\alpha}} e^{\gamma t}\
n_\mathrm{t}(\mathbf{r},t),\qquad t\gg c^{-1}.
\end{equation}
Substituting the last relation into the continuity equation
$$
\frac{\partial n(\mathbf{r},t)}{\partial t} +\mathrm{div}
\left(\mu \mathbf{E}\ n_\mathrm{c}(\mathbf{r},t)- D\ \nabla
n_\mathrm{c}(\mathbf{r},t)\frac{}{}\right)=0
$$
and taking into account the fact that most of carriers are
trapped, i. e. $n(\mathbf{r},t)\approx
n_\mathrm{t}(\mathbf{r},t)$, we arrive at
$$
\frac{\partial n(\mathbf{r},t)}{\partial t} +
$$
\begin{equation}\label{eq_transport_truncation}
+\mathrm{div}\left[e^{-\gamma t} {_0\textsf{D}_t^{1-\alpha}}
e^{\gamma t}\ \left( \mathbf{K}\  n(\mathbf{r},t)- C\ \nabla
n(\mathbf{r},t)\right)\right]=0,
\end{equation}
where $\textbf{K}=c^\alpha \tau_0\mu \textbf{E}$, $C=c^\alpha
\tau_0 D$ are coefficients, and $\mu,\ D,\ \mathbf{E}$ are a
mobility, a diffusion coefficient and a field intensity,
respectively. If $\alpha\rightarrow1$,
Eq.~(\ref{eq_transport_truncation}) becomes the standard
Fokker-Planck equation describing the normal transport. When
$\gamma=0$, the equation coincides with the dispersive transport
equation.

\section{Scale effect of transition from the dispersive regime to
the Gaussian one}

\begin{figure}[htp]
\centering
\includegraphics[width=0.45\textwidth]{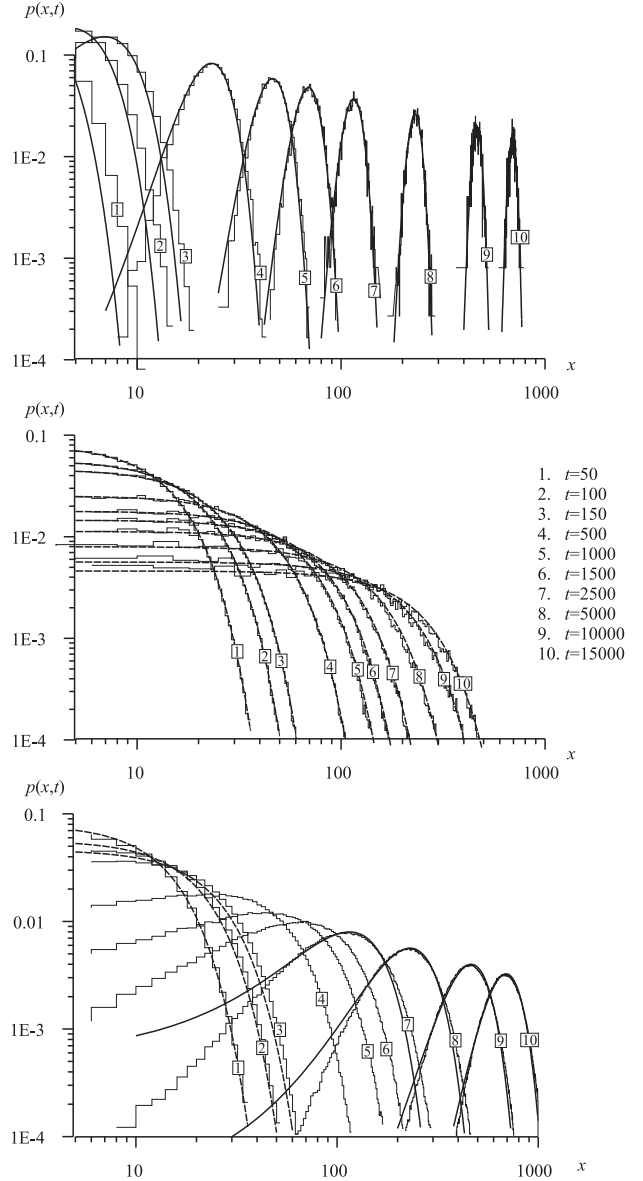}
\caption{The transition from the fractionally stable statistics to
the Gaussian one in the case of a one-sided random walk with
waiting times distributed according to law having truncated power
law tail. Solid lines represent Gaussian PDFs, dotted lines
correspond to fractionally stable densities.}\label{fig_truncated}
\end{figure}

Fig.~\ref{fig_truncated} represents numerically calculated spatial
distributions for one-sided random walks. With different waiting
time distributions, one can see the transition from fractionally
stable statistics to Gaussian statistics for waiting time
distribution having truncated power law tail. The upper graph
represents results for exponentially distributed times between
jumps. In the middle graph, waiting times are not truncated, the
exponent of the power law tail is equal to 0.5. In the lower
graph, waiting time distribution has the sharp truncation at time
$\tau_{\mathrm{tr}}=1500$ (arbitrary units). The exponential
distribution of waiting times has the same mathematical
expectation as truncated power law in the lower graph. In the case
of exponentially distributed waiting times, we see the fast
convergence to the Gaussian distribution, in the second case, the
distribution becomes fractionally stable law at some time $t$ and
maintains the form for all following times. In case of truncated
power law tails, the crossover between fractionally stable and
Gaussian statistics is observed.

Conduction current density at pulsed injection for the case of
exponentially truncated power-law waiting time distributions is
expressed as
$$
j(x,t)= e N
\exp\left[\frac{x}{l}{\left(\frac{\gamma}{c}\right)}^\alpha-\gamma
t \right]\times\ \ \ \qquad \ \qquad \ \qquad
$$
\begin{equation}\label{current_density_trunc}
\ \qquad \ \qquad\times c \left(\frac{x}{l}\right)^{-1/\alpha}
g^{(\alpha)}\left(c t \left(\frac{x}{l}\right)^{-1/\alpha}\right).
\end{equation}
Transient current density is found by substituting this expression
into the formula
$$
I(t)=\frac{1}{L}\int\limits_0^L j(x,t) dx.
$$

If $\alpha=1/2$, the expression for transient current takes the
form:
$$
I(t)=\frac{eN l\sqrt{c}}{L}\left\{ \frac{\exp(-\gamma
t)-\exp\left(-\left(\sqrt{\gamma
t}-\frac{1}{2\sqrt{\tau}}\right)^2\right)}{\sqrt{\pi t}}+\right.
$$
\begin{equation}\label{eq_current_truncation05}
\left.+\sqrt{\gamma} \left[\mathrm{erf}(\sqrt{\gamma
t})-\mathrm{erf}\left(\sqrt{\gamma
t}-\frac{1}{2\sqrt{\tau}}\right)\right] \right\}
\end{equation}

\begin{figure}[t]
\centering
\includegraphics[width=0.45\textwidth]{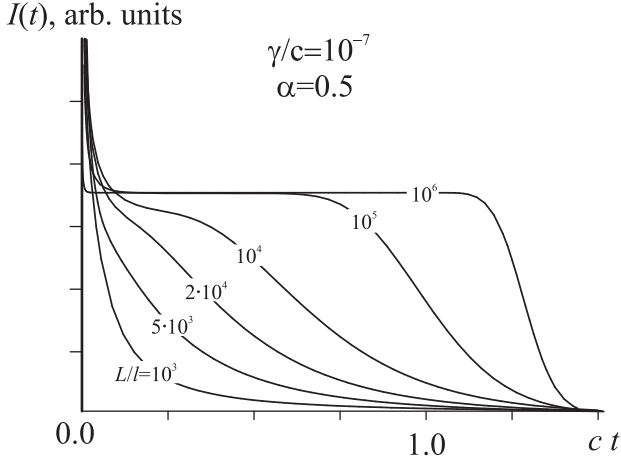}
\caption{Transient photocurrent in the case of truncated power law
distribution of waiting times for different values
of~$L/l$-ratio.}\label{tr_current_trunc}
\end{figure}

Fig.~\ref{tr_current_trunc} illustrates transformation of
transient current curves with increasing~$L/l$-ratio. When the
time of flight is much smaller than the truncation
time~$\gamma^{-1}$, the transport remains dispersive and does not
pass to Gaussian asymptotics. If $t_\mathrm{tr}$ is compared with
$\gamma^{-1}$ the shape of the transient current curves undergoes
modification, and they become inconsistent with the curves for
normal and dispersive transport. For $t_\mathrm{tr}\gg
\gamma^{-1}$, transport in the long-time asymptotic regime becomes
quasi-Gaussian.

\section{Frequency dependence of conductivity}

The frequency dependence of the real component of conductivity in
disordered semiconductors is usually fairly well described by the
power law:
\begin{equation}\label{sigma_omega}
\mathrm{Re}\ \sigma(\omega)=A \omega^\gamma,
\end{equation}
where the exponent $\gamma$ normally takes on values from 0.7 to 1
(\cite{Zvy:84}). The dependence of such type is characteristic of
a very broad class of materials.

Conductivity is related to mobility by the expression
$$
\sigma(\omega)=e \eta \mu(\omega).
$$
Here, $\eta$ is the concentration of effective carriers. The
Nyquist formula (generalized Einstein relation) linking mobility
with the diffusion coefficient at nonzero frequencies has the form
$$
\mu(\omega)=(e/kT)D(\omega),
$$
where the noise spectrum according the Wiener-Khintchin theorem is
expressed through the Fourier transform of the velocity
autocorrelation function
\begin{equation}\label{Weiner_Khinchin}
\mathrm{Re}\ D(\omega)=\int\limits_0^\infty \cos(\omega t)\langle
v(t) v(0)\rangle dt.
\end{equation}
This formula is important in that "a knowledge of the fluctuations
of the equilibrium ensemble in the absence of the electric field
permits a calculation of the linear response of the system
(mobility)" (\cite{Sch:73a}). Scher \& Lax showed that relation
(\ref{Weiner_Khinchin}) can be written out as:
\begin{equation}\label{D_omega1}
D(\omega)=-\frac{1}{6}\ \omega^2\int\limits_0^\infty dt\
e^{-i\omega t}\left\langle
[\mathbf{r}(t)-\mathbf{r}(0)]^2\right\rangle.
\end{equation}
The latter relation is possible to rewrite in the form
\begin{equation}\label{D_omega2}
D(\omega)=-\omega^2\int\limits_0^\infty dx\ x^2
\left[\widetilde{n}(x,s)\right]_{s=i\omega},
\end{equation}
where $\widetilde{n}(x,s)$ is the Laplace image of $n(x,t)$ with
respect to $t$.

Equation (\ref{eq_transport_truncation}) for one-dimensional case
without field assumes the form:
$$
\frac{\partial n(x,t)}{\partial t} = C\ e^{-\gamma t}
\frac{\partial^{1-\alpha}}{\partial t^{1-\alpha}} e^{\gamma t}\
\frac{\partial^2 n(x,t)}{\partial x^2} .
$$
The Laplace transform of this equation yields
$$
s\ \widetilde{n}(x,s)=C\ (s+\gamma)^{1-\alpha}\ \frac{\partial^2
\widetilde{n}(x,s)}{\partial x^2}+\delta(x).
$$
Substituting the solution of this equation,
$$
\widetilde{n}(x,s)=\frac{s^{-1/2}(s+\gamma)^{(\alpha-1)/2}}{\sqrt{C}}\exp\left(-\frac{|x|}{\sqrt{C}}\
\sqrt{s(s+\gamma)^{\alpha-1}}\right)
$$
into relation (\ref{D_omega2}) gives
$$
D(\omega)=2  C (\gamma+i\omega)^{1-\alpha}.
$$
Hence follows {\small
$$
\mathrm{Re}\ \sigma(\omega)=(e^2\eta/kT)\ \mathrm{Re}\ D(\omega)=
$$
\begin{equation}\label{sigma_theory}
=2 C(e^2\eta/kT)\
(\gamma^2+\omega^2)^{(1-\alpha)/2}\cos((1-\alpha)\arctan(\omega/\gamma)).
\end{equation}}
For frequencies $\omega\gg \gamma$, it is easily shown that
$$
\mathrm{Re}\ \sigma(\omega)=2 C(e^2\eta/kT)\
\omega^{1-\alpha}\sin(\pi\alpha/2).
$$

\begin{figure}[t]
\centering
\includegraphics[width=0.35\textwidth]{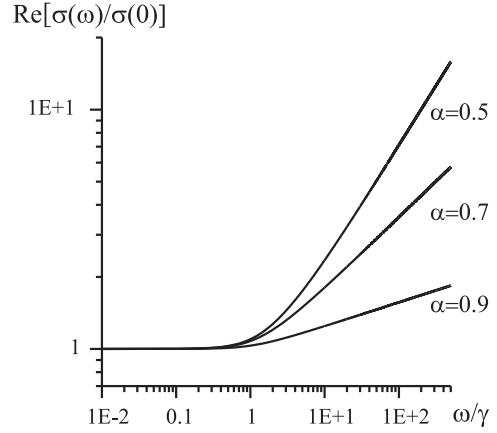}
\caption{Theoretical frequency dependencies of conductivity for
different~$\alpha$ values.}\label{cond_freq}
\end{figure}

Figure \ref{cond_freq} represents frequency-dependent conductivity
curves calculated from Eq.~(\ref{sigma_theory}). Formula
(\ref{sigma_theory}) predicts the power-law dependence of
conductivity on $\omega$ at high frequencies in the dispersive
transport case. The exponent may acquire values from 0 to 1; in
normal transport ($\alpha\rightarrow1$), conductivity is totally
frequency independent. In transport driven by the multiple
trapping mechanism, exponent $\alpha$ grows linearly with
temperature. Consequently, the exponent ${s=1-\alpha}$ in the
frequency dependence of conductivity in the case of alternating
current must linearly decrease with increasing temperature. Such
temperature behavior has been reported for a variety of
semiconductors (see, for instance, \cite{Gho:06}).

\section{Conclusion}

Here, transition from the dispersive transport regime to the
quasi-Gaussian one in disordered semiconductors is interpreted in
terms of truncated L\'evy distributions of waiting times. So,
polymer with Gaussian density of localized states is not exclusive
representative of materials that can show such behavior. The
phenomenon is more general and it is based on statistical rules
such as generalized limit theorem. Analytical results are
supported by numerical simulations.


\begin{thebibliography}{0}
\providecommand{\natexlab}[1]{#1}
\providecommand{\url}[1]{\texttt{#1}}
\providecommand{\urlprefix}{URL }
\expandafter\ifx\csname urlstyle\endcsname\relax
  \providecommand{\doi}[1]{doi:\discretionary{}{}{}#1}\else
  \providecommand{\doi}{doi:\discretionary{}{}{}\begingroup
  \urlstyle{rm}\Url}\fi

\end{thebibliography}


\begin{thebibliography}{40}
\providecommand{\natexlab}[1]{#1}
\providecommand{\url}[1]{\texttt{#1}} \expandafter\ifx\csname
urlstyle\endcsname\relax
  \providecommand{\doi}[1]{doi: #1}\else
  \providecommand{\doi}{doi: \begingroup \urlstyle{rm}\Url}\fi

\bibitem[Arkhipiov et al. (1983a)]{Ark:83} V. I. Arkhipov, A. I. Rudenko, A. M. Andriesh et al. \emph{Non-stationary injection currents in disordered
solids}. Kishinev,~1983.

\bibitem[Arkhipov et al. (1983b)]{Ark:83b} V. I. Arkhipov, Yu. A. Popova, A. I. Rudenko. \emph{Semiconductors} 17:1159,
1983.

\bibitem[Aroutiounian et al. (2000)]{Aro:00} V. M. Aroutiounian, M. Zh. Ghoolinian and H.
Tributsch. \newblock Fractal model of porous semiconductor.
\emph{Applied Surface Science} 162:122--132, 2000.

\bibitem[Barkai (2001)]{Bar:01} E. Barkai.  \newblock Fractional Fokker-Planck equation, solution, and application.
\newblock \emph{Phys. Rev. E}~63:046118, 2001.

\bibitem[B\"assler (1993)]{Bas:93} H. B\"assler. \newblock  Charge transport in disordered organic photoconductors. \newblock
\emph{Phys. Stat. Sol.} (b) 175:15--56,
1993.


\bibitem[Ghosh et al. (2006)]{Gho:06} P.~Ghosh, A.~Sarkar, A.~K.~Meikap, S.~K.~Chattopadhyay, S.~K.~Chatterjee,
M.~Ghosh. \newblock Electron transport properties of cobalt doped
polyaniline. \newblock \emph{J. Phys. D: Appl. Phys.}
39:3047--3052, 2006.

\bibitem[Koponen (1995)]{Kop:95} I.~Koponen. \newblock Analytic approach to the problem of convergence of
truncated L'evy flights towards the Gaussian stochastic process.
\newblock \emph{Phys. Rev. E} 52:1197-1199, 1995.

\bibitem[Madan \& Shaw (1988)]{Mad:88} A.~Madan, M.~P.~Shaw. \emph{The Physics and Applications
of Amorphous Semiconductors}. Academic Press Inc., Boston, 1988.

\bibitem[Mantegna \& Stanley (1994)]{Man:94} R.~N.~Mantegna and H.~E.~Stanley. \newblock Stochastic process with ultraslow
convergence to a Gaussian: The truncated L\'evy flight \newblock
\emph{Phys. Rev. Letters} 73:2946--2949, 1994.

\bibitem[Montroll \& Weiss (1965)]{Mon:65} E. W. Montroll, G. H.
Weiss. Random walks on lattices. \emph{J. Math. Phys.} 6:167--181,
1965.

\bibitem[Scher \& Lax (1973a)]{Sch:73a} H.~Scher and M.~Lax. \newblock Stochastic transport in a disordered solid. I.
Theory. \newblock \emph{Phys. Rev. B}~7:4491--4502, 1973.

\bibitem[Scher \& Lax (1973b)]{Sch:73b} H.~Scher and M.~Lax. \newblock Stochastic transport in a disordered solid. II.
Impurity conduction. \newblock \emph{Phys. Rev. B}~7:4502--4519,
1973.

\bibitem[Scher \& Montroll (1975)]{Sch:75} H.~Scher and E.~W.~Montroll. Anomalous transit-time dispersion in amorphous
solids. \emph{Phys. Rev. B}~12 (1975)~2455-2477.

\bibitem[Seki et al. (2003)]{Sek:03} K. Seki, M. Wojcik, M. Tachiya. Recombination kinetics in subdiffusive
media. \emph{J. Chem. Phys.} 119:7525--7533, 2003.

\bibitem[Seki et al. (2006)]{Sek:06} K. Seki, M. Wojcik, M. Tachiya. Dispersive-diffusion-controlled distance-dependent recombination
in amorphous semiconductors. \emph{J. Chem. Phys.} 124:044702,
2006.

\bibitem[Sibatov \& Uchaikin (2007)]{Sib:07} R. T. Sibatov, V. V.
Uchaikin. \newblock Fractional differential kinetics of charge
transport in unordered semiconductors. \newblock
\emph{Semiconductors} 41:335Ц-340, 2007.

\bibitem[Sibatov \& Uchaikin (2009)]{Sib:09} R. T. Sibatov, V. V. Uchaikin. \newblock Fractional differential approach to dispersive transport in
semiconductors. \newblock \emph{Physics Uspekhi} 52:1019Ц-1043,
2009.

\bibitem[Tiedje (1984)]{Tie:84} T. Tiedje. \newblock Investigations of charge transport in hydrogenated amorphous silicon. \newblock In:
J.~D.~Joannoulos  and G.~Lucovsky (Eds.), \emph{The Physics of
Hydrogenated Amorphous Silicon II. Electronic and Vibrational
Properties}. Springer-Verlag,  New York, 1984.

\bibitem[Tyutnev et al. (2005)]{Tyu:05} A. P. Tyutnev, V. S. Saenko, E. D. Pozhidaev, N. S. Kostyukov.
Dielektricheskie Svoistva Polimerov v Polyakh Ioniziruyushchikh
Izluchenii (Dielektriki i Radiatsiya, Kn. 5) [Dielectric
Properties of Polymers in Ionizing Radiation Fields (Dielectrics
and Radiation, Book 5)], Moscow: Nauka, 2005.

\bibitem[Uchaikin \& Zolotarev (1999)]{Uch:99a} V. V. Uchaikin and V. M. Zolotarev. \emph{Chance and
Stability.} VSP, Utrecht, tne Netherlands, 1999.

\bibitem[Uchaikin (1999)]{Uch:99b} V. V. Uchaikin. \newblock Subdiffusion and stable laws.
\newblock \emph{J. of Exper. and Theor. Phys.} 115:2113--2132, 1999.

\bibitem[Uchaikin \& Sibatov (2005)]{Uch:05} V. V. Uchaikin and R. T. Sibatov.  \newblock Fractional derrivatives in theory of
semiconductors.  \newblock \emph{Surveys of Appl. and Industr.
Math.} 12:540--542, 2005.

\bibitem[Uchaikin \& Sibatov (2008)]{Uch:08} V. V. Uchaikin and R. T. Sibatov. \newblock  Fractional theory for
transport in disordered semiconductors. \newblock
\emph{Communications in Nonlinear Science and Numerical
Simulation} 13:715Ц727, 2008.

\bibitem[Zvyagin (1984)]{Zvy:84} I. P. Zvyagin. \newblock \emph{Kineticheskie Yavleniya v
Neuporyadochennykh Poluprovodnikakh} (\emph{Kinetic Phenomena in
Disordered Semiconductors}). Moscow: Izd. MGU, 1984.

\end{thebibliography}
\end{document}